\documentclass[useAMS,usenatbib,usegraphicx]{mn2e}

\newcommand{\porb}{\mbox{$P_{\mathrm{orb}}$}}
\newcommand{\system}{\mbox{SDSS1210}}

\newcommand{\twd}{\mbox{$T_{\mathrm{eff,WD}}$}}
\newcommand{\tsec}{\mbox{$T_{\mathrm{sec}}$}}
\newcommand{\Msun}{\mbox{$\mathrm{M}_{\odot}$}}
\newcommand{\Rsun}{\mbox{$\mathrm{R}_{\odot}$}}
\newcommand{\mwd}{\mbox{$M_{\mathrm{WD}}$}}
\newcommand{\rwds}{\mbox{$R_{\mathrm{WD}}$}}
\newcommand{\rwda}{\mbox{$r_{\mathrm{WD}}$}}
\newcommand{\kwd}{\mbox{$K_{\mathrm{WD}}$}}
\newcommand{\kwdm}{\mbox{$K^{m}_{\mathrm{WD}}$}}
\newcommand{\gamwd}{\mbox{$\gamma_{\mathrm{WD}}$}}
\newcommand{\msec}{\mbox{$M_{\mathrm{sec}}$}}
\newcommand{\rsecs}{\mbox{$R_{\mathrm{sec}}$}}
\newcommand{\rsecv}{\mbox{$R_{\mathrm{sec,vol.aver.}}$}}
\newcommand{\rseca}{\mbox{$r_{\mathrm{sec}}$}}
\newcommand{\ksec}{\mbox{$K_{\mathrm{sec}}$}}
\newcommand{\ksecm}{\mbox{$K^{m}_{\mathrm{sec}}$}}
\newcommand{\gamsec}{\mbox{$\gamma_{\mathrm{sec}}$}}
\newcommand{\Lines}[3]{\Ion{#1}{#2}\,$\lambda\lambda$\,#3}
\newcommand{\Ion}[2]{#1{\,\scriptsize #2}}
\newcommand{\kms}{\mbox{$\mathrm{km\,s^{-1}}$}}
\newcommand{\zspec}{\mbox{$\mathrm{z_{WD,spec}}$}}
\newcommand{\zphot}{\mbox{$\mathrm{z_{WD}}$}}
\newcommand{\vscale}{\mbox{$V_{\mathrm{S}}$}}
\usepackage{amssymb}
\title[SDSS\,1210+3347]{Post Common Envelope Binaries from SDSS. XV: Accurate stellar parameters for a cool $0.4\,\Msun$ white dwarf and a $0.16\,\Msun$ M-dwarf in a 3\,h eclipsing binary}

\author[S.Pyrzas et al.]{
S. Pyrzas$^{1}$\thanks{E-mail: S.Pyrzas@warwick.ac.uk},
B. T. G\"ansicke$^{1}$, 
S. Brady$^{2}$,
S. G. Parsons$^{1}$,
T. R. Marsh$^{1}$,
D. Koester$^{3}$,\newauthor
E. Breedt$^{1}$, 
C. M. Copperwheat$^{1}$, 
A. Nebot G\'omez-Mor\'an$^{4}$, 
A. Rebassa-Mansergas$^{5}$, \newauthor 
M. R. Schreiber$^{5}$ and
M. Zorotovic$^{5}$\\
$^{1}$Department of Physics, University of Warwick, Coventry, CV4 7AL, UK\\ 
$^{2}$AAVSO, 5 Melba Drive, Hudson, NH 03051, USA\\
$^{3}$Institut f\"ur Theoretische Physik und Astrophysik, University of Kiel, 24098 Kiel, Germany\\
$^{4}$Universit\'e de Strasbourg, CNRS, UMR7550, Observatoire Astronomique de Strasbourg, 11 Rue de l'Universit\'e, F-67000, Strasbourg, France\\
$^{5}$Departamento de F\'isica y Astronom\'ia, Facultad de Ciencias, Universidad de Valpara\'iso, Avenida Gran Bretana  1111, Valpara\'iso, Chile\\ 
}
\begin{document}

\date{Accepted 2011 August 31.  Received 2011 August 26; in original form 2011 July 27}

\pagerange{\pageref{firstpage}--\pageref{lastpage}} \pubyear{2011}

\maketitle

\label{firstpage}

\begin{abstract}
We identify SDSSJ\,121010.1+334722.9 as an eclipsing post-common-envelope binary, with an orbital period of $\porb\,=\,2.988\,\mathrm{h}$, containing a very cool, low-mass, DAZ white dwarf and 
a low-mass main-sequence star of spectral type M5. A model atmosphere analysis of the metal absorption lines detected in the blue part of the optical spectrum, along with the GALEX 
near-ultraviolet flux, yields a white dwarf temperature of $\twd\,=\,6000\,\pm\,200$\,K and a metallicity value of $\log[\mathrm{Z/H}]\,=\,-2.0\,\pm\,0.3$.  The \Lines{Na}{I}{8183.27,8194.81} 
absorption doublet is used to measure the radial velocity of the secondary star, $\ksec\,=\,251.7\,\pm\,2.0\,\kms$ and \Ion{Fe}{I} absorption lines in the blue part of the spectrum provide the 
radial velocity of the white dwarf, $\kwd\,=\,95.3\,\pm\,2.1\,\kms$, yielding a mass ratio of $q\,=\,0.379\,\pm\,0.009$. Light curve model fitting, using the Markov Chain Monte Carlo (MCMC) 
method, gives the inclination angle as $i\,=\,(79.05^{\circ}\,-\,79.36^{\circ})\,\pm\,0.15^{\circ}$, and the stellar masses as $\mwd\,=\,0.415\,\pm\,0.010\,\Msun$ and 
$\msec\,=\,0.158\,\pm\,0.006\,\Msun$. Systematic uncertainties in the absolute calibration of the photometric data influence the determination of the stellar radii. The radius of the white 
dwarf is found to be $\rwds\,=\,(0.0157\,-\,0.0161)\,\pm\,0.0003\,\Rsun$ and the volume-averaged radius of the tidally distorted secondary is $\rsecv\,=\,(0.197\,-\,0.203)\,\pm\,0.003\,\Rsun$. 
The white dwarf in \system\, is a very strong He-core candidate.
\end{abstract}



\section{Introduction}
\label{sec:intro}

Our understanding of stellar structure and evolution leads to the fundamental prediction that the masses and radii of stars obey certain mass-radius (M-R) relations. The calibration and 
testing of the M-R relations requires accurate and model-independent measurements of stellar masses and radii, commonly achieved with eclipsing binaries 
\citep[e.g.][]{andersen91-1,southworth+clausen07-1}.

Among main-sequence (MS) stars, M-dwarfs of low mass ($<\,0.3\Msun$), are the most ubiquitous. However, few eclipsing low-mass MS+MS binaries are known (e.g. \citealt{lopez-morales07-1,
moralesetal09-1,cakirli+ibanoglu10-1,irwinetal10-1,dimitrov10-1} and references therein) and have accurate measurements of their masses and radii, affecting the calibration of the low-mass end 
of the MS M-R relation. To further complicate matters, existing measurements consistently result in radii up to 15\% larger and effective temperatures 400\,K or more below the values predicted 
by theory \citep[e.g.][]{ribas06-1,lopez-morales07-1}. This is not only the case for low-mass MS+MS binaries \citep{bayless+orosz06-1}, but it is also present in field stars 
\citep{bergeretal06-1,moralesetal08-1} and the host stars of transiting extra-solar planets \citep{torres07-1}.

The situation is similar for white dwarfs, the most common type of stellar remnant. Very few white dwarfs have model-independent measurements of their masses and radii 
\citep[see][]{parsonsetal10-1}, and eclipsing WD+WD binaries have only recently been discovered \citep{steifadtetal10-1,parsonsetal11-1,brownetal11-1}. Consequently, the finite temperature 
M-R relation of white dwarfs \citep[e.g.][]{wood95-1,paneietal00-2} remains largely untested by observations \citep{provencaletal98-1}. 

An alternative approach leading to accurate mass and radius measurements for WDs and MS stars is the study of eclipsing WD+MS binaries. Until recently, the population of eclipsing WD+MS 
binaries had stagnated with only seven systems known \citep[see][for a list]{pyrzasetal09-1}, a direct result of the small number of the entire WD+MS binaries sample 
\citep[$\sim\,30$ systems; ][]{schreiber+gaensicke03-1}. 

However, in recent years, progress has been made thanks to the Sloan Digital Sky Survey \citep[SDSS; ][]{yorketal00-1}. A dedicated search for WD+MS binaries contained in the spectroscopic
SDSS Data Release 6 \citep{adelmanetal08-1} and DR7 \citep{abazajianetal09-1} yielded more than 1600 systems \citep[e.g.][]{rebassamansergasetal10-1}, of which $\sim\,1/3$ are (short-period) 
post-common-envelope binaries (PCEBs) \citep{schreiberetal08-1}. The majority of these PCEBs contain low-mass, late-type M dwarfs \citep{rebassamansergasetal10-1}, while a large percentage of 
the WD primaries are of low-mass as well \citep{rebassamansergasetal11-1}. 

A significant fraction of eclipsing systems should exist among this sample of PCEBs. Identifying and studying these eclipsing systems will substantially increase the observational constraints 
on the M-R relation of both WDs and MS stars. Therefore, we have begun the first dedicated search for eclipsing WD+MS binaries in the SDSS, and 5 new systems have already been published 
(\citealt{nebotgomezmoranetal09-1,pyrzasetal09-1}, but see also \citealt{drakeetal10-1} for a complementary sample).

SDSSJ\,121010.1+334722.9 (henceforth \system), the subject of this paper, is one of the new systems identified in this search. In what follows, we present our observations 
(Sec.\,\ref{sec:obsnred}), determine the orbital period and ephemeris (Sec.\,\ref{sec:porbephem}) and analyse the spectrum of the white dwarf (Sec.\,\ref{sec:wdanalysis}). Radial velocity 
measurements (Sec.\,\ref{sec:radvel}) combined with light curve fitting (Sec.\,\ref{sec:fitting}) lead to the determination of the masses and radii of the binary components 
(Sec.\,\ref{sec:result}). We also explore the past and future evolution of the system (Sec.\,\ref{sec:evolution}).


\section{Target infromation, observations and reductions}
\label{sec:obsnred}

\system\, was discovered by \citet{rebassamansergasetal10-1} as a WDMS binary dominated by the flux of a low-mass companion with a spectral type M5V, suggesting that the white dwarf must be 
very cool. Inspecting the \Lines{Na}{I}{8183.27,8194.81} doublet in the six SDSS sub-spectra\footnote{The sub-exposures that are co-added to produce one SDSS spectrum of a given object} with 
exposure times of 15--30\,min taken over the course of three nights, we found large radial velocity variations that strongly suggested an orbital period of a few hours. We obtained time-series 
photometry of \system\, with a 16-inch telescope equipped with an ST8-XME CCD camera, with the aim to measure the orbital period from the expected ellipsoidal modulation, and immediately 
detected a shallow eclipse in the light curve. Enticed by this discovery, we scheduled \system\, for additional high-time resolution photometry, using RISE on the Liverpool Telescope (LT), 
with which a total of 9 eclipses were observed.

Table\,\ref{tab:coordmag} lists the SDSS coordinates and magnitudes of \system\, and the three comparison stars used in the analysis presented in this paper, while Table\,\ref{tab:obslog}
summarises our photometric and spectroscopic observations. We note that \system\, has a GALEX \citep{morrisseyetal07-1} near-ultraviolet (NUV) detection, but no far-ultraviolet (FUV) detection.

\begin{table}
\setlength{\tabcolsep}{0.7ex}
\centering
\caption{SDSS coordinates and $u,g,r,i,z$ magnitudes of the target \system\, and the comparison stars used in the analysis. We also provide the GALEX near-UV magnitude of \system.}
\label{tab:coordmag}
\begin{tabular}{@{}ccccccccc@{}}
\hline 
Star & RA & Dec & $u$ & $g$ & $r$ & $i$ & $z$ & NUV \\ 
\hline 
T  & 182.54221 & 33.78969 & 18.10 & 16.94 & 16.16 & 14.92 & 14.02 & 20.821  \\
C1 & 182.55470 & 33.76832 & 17.72 & 15.98 & 15.33 & 15.11 & 15.02 &   \\
C2 & 182.54229 & 33.73406 & 19.95 & 17.33 & 16.01 & 15.33 & 14.94  & \\
C3 & 182.62616 & 33.78141 & 16.85 & 15.80 & 15.46 & 15.34 & 15.34   & \\
\hline
\end{tabular}
\end{table}

\begin{table}
\setlength{\tabcolsep}{0.6ex}
\centering
\caption{Log of the photometric and spectroscopic observations. For the LT observations, we also provide the number of one-hour observing blocks per night.}
\label{tab:obslog}
\begin{tabular}{@{}ccccccc@{}}
\hline
Date        & Telescope & Filter/Grating & Exp.\,[s] & Blocks & Frames & Eclipses \\ 
\hline
2009 Apr 01 &    LT     &       V+R      &     5     &   1    &  708   &   1      \\
2009 Apr 02 &    LT     &       V+R      &     5     &   2    & 1416   &   0      \\
2009 Apr 03 &    LT     &       V+R      &     5     &   2    & 1416   &   1      \\
2009 Apr 04 &    LT     &       V+R      &     5     &   2    & 1416   &   1      \\
2009 Apr 05 &    LT     &       V+R      &     5     &   3    & 2124   &   1      \\
2009 Apr 06 &    LT     &       V+R      &     5     &   1    &  708   &   0      \\
2009 Apr 29 &   WHT     &  R600B/R1200R  &   900     &   -    &    1   &   -      \\
2009 May 02 &   WHT     &  R600B/R1200R  &   900     &   -    &    3   &   -      \\
2010 Apr 23 &   WHT     &  R600B/R1200R  &   600     &   -    &    1   &   -      \\
2010 May 18 &   WHT     &  R600B/R600R   &   900     &   -    &   12   &   -      \\
2011 Feb 06 &    LT     &       V+R      &     5     &   1    &  720   &   1      \\ 
2011 Mar 02 &    LT     &       V+R      &     5     &   1    &  720   &   1      \\ 
2011 Apr 02 &    LT     &       V+R      &     5     &   1    &  720   &   1      \\
2011 May 08 &    LT     &       V+R      &     5     &   1    &  720   &   1      \\
2011 Jul 03 &    LT     &       V+R      &     5     &   1    &  720   &   1      \\
\hline
\end{tabular}
\end{table}

\subsection{Photometry: LT/RISE}
\label{subsec:photom}

Photometric observations were obtained with the robotic 2.0\,m Liverpool Telescope (LT) on La Palma, Canary Islands, using the high-speed frame-transfer CCD camera RISE \citep{steeleetal04-1}
equipped with a single wideband V+R filter \citep{steeleetal08-1}. Observations were carried out in one hour blocks, using a 2\,x\,2 binning mode with exposure times of 5 seconds.

The data were de-biased and flat-fielded in the standard fashion within the LT reduction pipeline and aperture photometry was performed using {\sc sextractor} \citep{bertin+arnouts96-1} in
the manner described in \citet{gaensickeetal04-1}.

\begin{figure}
\centering
 \includegraphics[angle=-90,width=84mm]{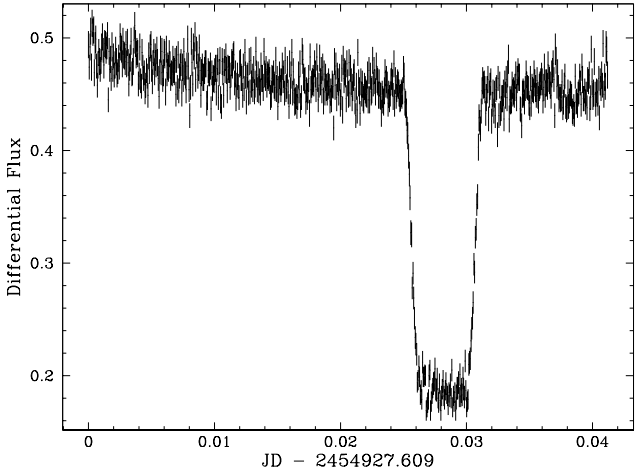}
  \caption{Sample light curve of \system\, obtained with a 5\,sec. cadence using RISE on the LT on April 05, 2009.}
  \label{fig:samplelc}
\end{figure}

A sample light curve is shown in Figure\,\ref{fig:samplelc}. The out-of-eclipse variation is ellipsoidal modulation, arising from the tidally deformed secondary.

\subsection{Spectroscopy: WHT/ISIS}
\label{subsec:spectra}

Time-resolved spectroscopy was carried out at the 4.2\,m William Herschel Telescope (WHT) on La Palma, Canary Islands, equipped with the double-armed Intermediate dispersion Spectrograph 
and Imaging System (ISIS). The spectrograph was used with a 1\arcsec\, slit, and an 600 lines/mm grating (R600B/R600R) on each of the blue and red arms, although a few spectra were obtained 
with a 1200 lines/mm grating on the red arm (R1200R). Both the EEV12 CCD on the blue arm and the REDPLUS CCD on the red arm were binned by three in the spatial direction and  two in the 
spectral direction. This setup resulted in an average dispersion of 0.88\AA\, per binned pixel over the wavelength range $3643 - 5137$\AA\, (blue arm) and 0.99\AA\, per binned pixel over the 
wavelength range $7691 - 9184$\AA\ (red arm, R600R). From measurements of the full width at half maximum of arclines and strong skylines, we determine the resolution to be 1.4\AA.

The spectra were reduced using the {\sc starlink}\footnote{Maintained and developed by the Joint Astronomy Centre and available from http://starlink.jach.hawaii.edu/starlink} packages 
{\sc kappa} and {\sc figaro} and then optimally extracted \citep{horne86-1} using the {\sc pamela}\footnote{Available from http://www.warwick.ac.uk/go/trmarsh} code \citep{marsh89-1}. The 
wavelength scale was derived from Copper-Neon and Copper-Argon arc lamp exposures taken every hour during the observations, which we interpolated to the middle of each of the science exposures. 
For the blue arm the calibration was determined from a 5th order polynomial fit to 25 lines, with a root mean square (RMS) of 0.029\AA. The red arm was also fitted with a 5th order polynomial, 
to 17 arclines. The RMS was 0.032\AA.


\section{Orbital period and ephemeris}
\label{sec:porbephem}

We determined the orbital period and ephemeris of \system\, through mid-eclipse timings. This was achieved as follows: 

Mid-eclipse times were measured by mirroring the observed eclipse profile around an estimate of the eclipse centre and shifting the mirrored profile against the original until the best overlap 
was found. This method is particularly well-suited for the box-shaped eclipse profiles in (deeply) eclipsing PCEBs. 

An initial estimate of the cycle count was then obtained by fitting eclipse phases $(\phi^{\rm{observed}}_{\rm{0}} - \phi^{\rm{fit}}_{\rm{0}})^{-2}$ over a wide range of trial periods. Once an 
unambiguous cycle count was established, a linear fit, of the form $T\,=\,T_0\,+\,\porb\times E$, was performed to the times of mid-eclipse versus cycle count, yielding a preliminary orbital
ephemeris.

Subsequently, we phase-folded our data set using this preliminary ephemeris and proceeded with the light curve model fitting (see Sec.\,\ref{sec:fitting}). Having an accurate model at hand, 
we re-fitted each light curve individually. This provides a robust estimate of the error on the mid-eclipse time, as our code includes the time of mid-eclipse $T_0$ as a free parameter.

Repeating the cycle count determination and the linear ephemeris fitting, as described above, we obtain the following ephemeris for \system, 

\begin{equation}
\label{eq:ephe}
\mathrm{MJD}\left(\mathrm{BTDB}\right)\,=\,54923.033\,686(6)\,+\,0.124\,489\,764(1)\,E
\end{equation} 
\noindent
calculated on a Modified Julian Date-timescale and corrected to the solar system barycentre, with the numbers in parentheses indicating the error on the last digit. Thus, \system\, has an 
orbital period of $\porb\,=\,2.987\,754\,336(24)\,\mathrm{h}$. The mid-eclipse times, the observed minus calculated values (O-C) and their respective errors are given in 
Table\,\ref{tab:midecltim}. Given the short baseline, there is as yet no evidence for period changes which are frequently seen in such binaries \citep[e.g.][]{parsonsetal10-2}.

\begin{table}
\setlength{\tabcolsep}{0.8ex}
\centering
\caption{Times of mid-eclipse (and their errors), O-C values (and their errors) and cycle number for the ephemeris of \system. Mid-eclipse times and errors are in MJD(BTDB), O-C values and 
errors are in seconds.}
\label{tab:midecltim}
\begin{tabular}{@{}ccccl@{}}
\hline
Mid-Eclipse [d] & Error [d] & O-C [s] & Error [s] & Cycle \\ 
\hline
54923.0336744 & 0.0000060 & -1 & 1 &    0 \\
54925.0255324 & 0.0000082 &  1 & 1 &   16 \\
54926.1459281 & 0.0000069 & -0 & 1 &   25 \\
54927.1418460 & 0.0000087 & -0 & 1 &   33 \\
55599.1376175 & 0.0000061 &  3 & 1 & 5431 \\
55623.0396100 & 0.0000056 & -1 & 1 & 5623 \\
55654.0375754 & 0.0000081 & -0 & 1 & 5872 \\
55690.0151216 & 0.0000063 &  1 & 1 & 6161 \\
55745.9109933 & 0.0000069 & -2 & 1 & 6610 \\
\hline
\end{tabular}
\end{table}


\section{Spectroscopic analysis}
\label{sec:wdanalysis}

\begin{figure}
\includegraphics[angle=-90,width=\columnwidth]{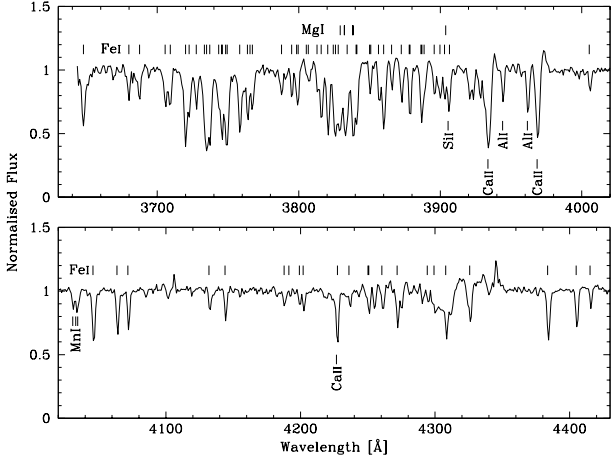}
\caption{\label{f-lineid}The normalised average WHT spectrum in the
  white dwarf restframe, along with line identifications for
  absorption lines originating in the white dwarf photosphere.}
\end{figure}

Whereas the SDSS spectrum of \system\, remained inconclusive with respect to the nature of the white dwarf \citep{rebassamansergasetal10-1}, our blue-arm WHT spectroscopy immediately revealed 
a host of narrow metal lines that exhibit radial velocity variations anti-phased with respect to those of the M-dwarf. The WHT spectra obtained in May 2010, averaged in the white dwarf 
restframe and continuum-normalised, are shown in Fig.\,\ref{f-lineid} and illustrate the wealth of absorption lines from Mg, Al, Si, Ca, Mn, and Fe. Similar metal lines have been detected in
the optical spectra of a few other cool PCEBs, e.g. RR\,Cae \citep{zuckermanetal03-1} or LTT\,560 \citep{tappertetal07-1}, and indicate accretion of mass via a wind from the M-dwarf.

\begin{figure}
\includegraphics[angle=-90,width=\columnwidth]{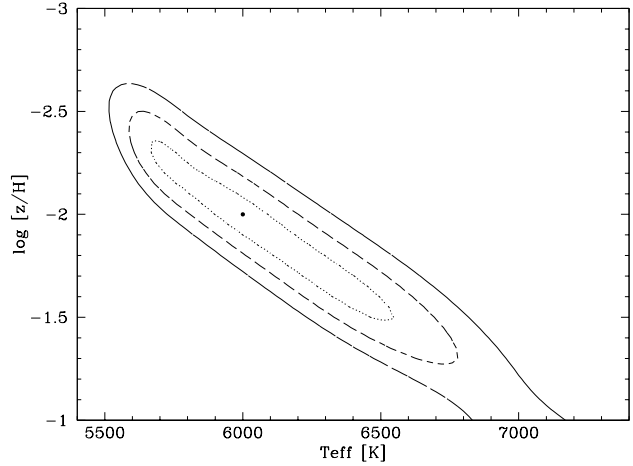}
\caption{\label{f-chi2} Results of model spectra fitting to the average WHT spectrum. The single, big point indicates the best fit solution. The contours indicate the regions where the $\chi^2$ of
the fit is within 1, 2 and 3$\,\sigma$ (dotted, short-dashed and long-dashed lines respectively) of the minimum (single point).}
\end{figure}

We have analysed the blue WHT spectra using hydrogen-dominated but metal-polluted (DAZ) spectra calculated with the stellar atmosphere code described by \citet{koester10-1}. We fixed the 
surface gravity to $\log g=7.70$, as determined from the fits to the LT light curve (Sect.\,\ref{sec:fitting}). The model grid covered effective temperatures 
$5400\,\mathrm{K}\,\le\,\twd\,\le\,7400\,\mathrm{K}$ in steps of 200\,K and metal and He abundances of $\log[\mathrm{Z/H}]\,=\,-3.0, -2.3, -2.0, -1.3, -1.0$, with all relevant elements up to 
zinc included, and fixed their relative abundances ratios to the respective solar values. We then fitted the model spectra to the average WHT spectrum in the range 3645--3930\,\AA, 
where the contribution of the M-dwarf is entirely negligible. A good fit is found for $\twd\,\simeq\,6000$\,K and metal abundances at $\simeq\,0.01$ their solar values, however, the effective 
temperature and the metal abundances are strongly correlated (Fig.\,\ref{f-chi2}). 

\begin{figure*}
\includegraphics[angle=-90,width=\textwidth]{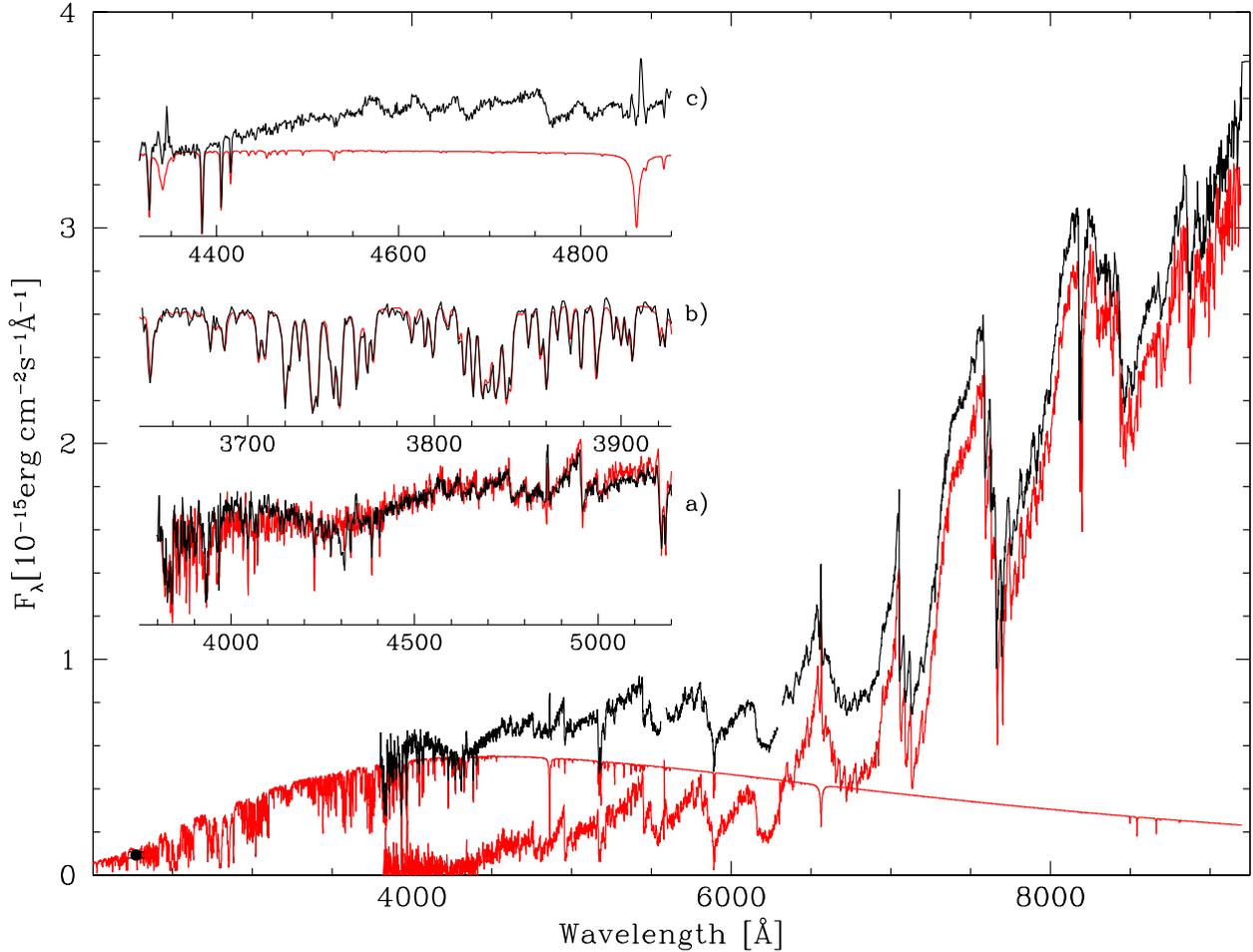}
\caption{Spectral modelling of SDSS1210. Main panel: the SDSS spectrum (black) and the GALEX NUV flux (black point), along with the best-fit white dwarf model (red$^{*}$, $\twd\,=\,6000$\,K, 
$\log g=7.70$, $\log[\mathrm{Z/H}]\,=\,-2.0$) and the best-fit M-dwarf template for the companion (red$^{*}$, spectral type M5). Inset (a): The sum of the white dwarf model and M-dwarf 
template provide a good match to the blue end of the SDSS spectrum (black), with the low flux of the M-dwarf template being the dominant limitation. Inset (b): Best-fit white dwarf model 
(red$^{*}$) and the average WHT spectrum for $\lambda<3930$\,\AA, where the M-dwarf contributes practically nothing to the observed flux. Inset (c): Best-fit white dwarf model (red$^{*}$) and 
the average WHT spectrum (black) illustrating the weakness of the H$\beta$ and H$\gamma$ lines of the white dwarf. Increasing the temperature very rapidly results in Balmer lines and/or a NUV 
flux that are inconsistent with the observations. $^{*}$The coloured Figure is available in the online version only.}
\label{fig:specfit}
\end{figure*}

This degeneracy is lifted by including the GALEX detection of \system, as the predicted near-UV flux is a strong function of the effective temperature. The uncertainty in the absolute flux 
calibrations of our WHT spectra and the GALEX observations introduces a small systematic uncertainty on the final result, and we settle for $\twd\,=\,6000\,\pm\,200$\,K and 
$\log[\mathrm{Z/H}]\,=\,-2.0\,\pm\,0.3$ . Independently, the weakness of the Balmer lines in the WHT spectrum also requires that $\twd\,\lesssim\,6400\,\rm{K}$. The spectral modelling of
\system\, is illustrated on Fig.\,\ref{fig:specfit}.

Adopting the white dwarf radius from the light curve fit (Sect.\,\ref{sec:fitting} and \ref{sec:result}), $\rwds\,=\,0.0159\,\Rsun$, the flux-scaling factor of the best-fit spectral model 
implies a distance of $d\,\simeq\,50\,\pm\,5$\,pc, which is in good agreement with $d\,\sim\,66\,\pm\,34$\,pc estimated by \citet{rebassamansergasetal10-1} from fitting the M-dwarf.

The detection of metals in the photosphere of the white dwarf allows an estimate of the accretion rate \citep[e.g.][]{dupuisetal93-1,koester+wilken06-1}, as long as the system is in 
accretion-diffusion equilibrium. In cool, hydrogen-rich atmospheres, such as the one in \system, the diffusion time scales of the different metals detected in the WHT spectrum vary by a factor 
of $\sim\,2$ for a given temperature, and are, for $\twd\,=\,6000\,\rm{K}$, in the range 30000-60000\,years\footnote{For completeness, we note that because we have adopted solar abundance 
ratios for the metals, these small differences in diffusion time scales imply slightly non-solar ratios in the accreted material. In principle, the individual metal-to-metal ratios can be 
determined from the observed spectrum of the white dwarf, and hence allow to infer the abundances of the companion star, however, this requires data with substantially higher spectral 
resolution to resolve the line blends.}. It is plausible to assume that the average accretion rate over the diffusion time scales involved is constant, as the binary configuration (separation 
of the two stars, Roche-lobe filling factor of the companion) changes on much longer time scales. Summing up the mass fluxes at the bottom of the convective envelope, and taking into account 
the uncertainties in \twd\ and the metal abundances, gives $\dot M\,\simeq\,(5\,\pm\,2)\times10^{-15}\,\mathrm{\Msun\,yr^{-1}}$. There are now three PCEBs with similar stellar components that 
have measured accretion rates, RR\,Cae ($\dot M\,\simeq\,4\times10^{-16}\,\mathrm{\Msun\,yr^{-1}}$; \citealt{debes06-1}), LTT\,560 ($\dot M\,\simeq\,5\times10^{-15}\,\mathrm{\Msun\,yr^{-1}}$; 
\citealt{tappertetal11-2}), and \system\, ($\dot M\,\simeq\,5\times10^{-15}\,\mathrm{\Msun\,yr^{-1}}$).

Whereas \system\, and LTT\,560 have similar orbital periods, the period of RR\,Cae is roughly twice as long, suggesting that the efficiency of wind-accretion decreases as the binary separation 
and Roche-lobe size of the companion increase, as is expected. A more systematic analysis of the wind-loss rates of M-dwarfs and the efficiency of wind accretion in close binaries would be 
desirable, but will require a much larger sample of systems.


\section{The spectroscopic orbit}
\label{sec:radvel}

Radial velocities of the binary components have been measured from the \Lines{Fe}{I}{4045.813,4063.594,4071.737,4132.058,4143.869} absorption lines for the white dwarf and the 
\Lines{Na}{I}{8183.27,8194.81} absorption doublet for the secondary star.

The \Ion{Fe}{I} lines were simultaneously fitted with a second-order polynomial plus five Gaussians of common width and a separation fixed to the corresponding laboratory values. A sine fit to 
the radial velocities, phase-folded using the orbital ephemeris (Equation\,\ref{eq:ephe}) yields $\kwd\,=\,95.3\,\pm\,2.1\,\kms$ and $\gamwd\,=\,24.2\,\pm\,1.4\,\kms$.

The \Ion{Na}{I} doublet was fitted with a second-order polynomial plus two Gaussians of common width and a separation fixed to the corresponding laboratory value. A sine fit to the radial
velocities, phase-folded using the orbital ephemeris yields $\ksec\,=\,251.7\,\pm\,2.0\,\kms$ and $\gamsec\,=\,12.2\,\pm\,0.9\,\kms$.

\begin{figure}
\centering
 \includegraphics[angle=-90,width=84mm]{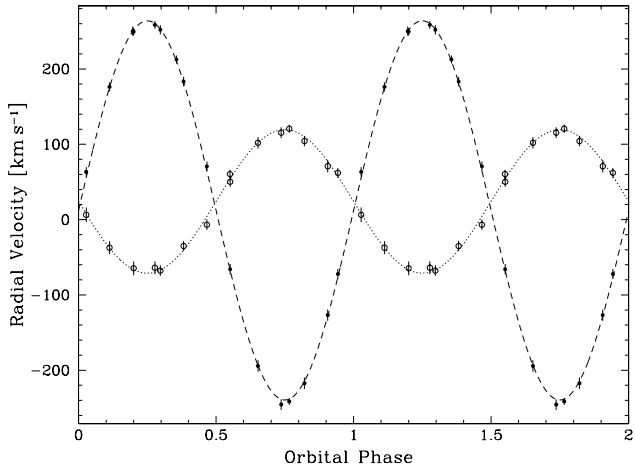}
  \caption{Phase-folded radial velocity curves of the secondary star (filled circles) and the white dwarf (open circles), with their respective errors. Also shown are the sine fits to the
velocities of both components. A full cycle is repeated for clarity.}
  \label{fig:rvorb}
\end{figure}

Figure\,\ref{fig:rvorb} shows the measured radial velocities phase-folded on the orbital period and the corresponding sine-fits.

Knowledge of both radial velocities allows us to obtain the mass ratio $q$ of the binary, namely $q\,=\,\kwd/\ksec\,=\,0.379\,\pm\,0.009$. We tentatively interpret the difference between
\gamwd\, and \gamsec\, as the gravitational redshift of the white dwarf \zspec, which yields $\zspec\,=\,11.9\,\pm\,1.7\,\kms$ (see also Sec.\,\ref{sec:result}).

\section{Light curve modelling}
\label{sec:fitting}

To obtain the stellar parameters of the binary components, light curve models were fitted to the data using {\sc lcurve} (see \citealt{copperwheatetal10-1} for a description, 
as well as \citealt{pyrzasetal09-1}; \citealt{parsonsetal10-1}; \citealt{parsonsetal11-1} for further applications).

\subsection{Code input}

The code computes a model based on input system parameters supplied by the user. The physical parameters defining the models are (i) the mass ratio $q = \msec/\mwd$, (ii) the binary
inclination $i$, (iii) the stellar radii scaled by the binary separation $\rwda\,=\,\rwds/a$ and $\rseca\,=\,\rsecs/a$, (iv) the unirradiated stellar temperatures \twd\ and \tsec, (v) the sum 
of the unprojected stellar orbital speeds $\vscale\,=\,\left(\kwd\,+\,\ksec\right)/\mathrm{sin}\,i$, (vi) the time of mid-eclipse of the white dwarf $T_0$, (vii) limb- and gravity darkening 
coefficients and (viii) the distance $d$. The code accounts for the distance simply as a scaling factor that can be calculated very rapidly for any given model, and so it does not enter the 
optimisation process. All other parameters can be allowed to vary during the fit. 

\subsection{Free and fixed parameters}

During the minimisation, we kept \twd\, fixed at $\twd\,=\,6000\,\mathrm{K}$. The gravity darkening of the secondary was also kept fixed at $0.08$ (the usual value for a convective atmosphere). 
Limb darkening coefficients were also held fixed. For the white dwarf we calculated quadratic limb darkening coefficients from a white dwarf model with $\twd\,=\,6000\,\mathrm{K}$ and 
$\mathrm{log}\,g\,=\,7.70$, folded through the RISE filter profile. The corresponding values were found to be a$\,=\,0.174$ and $b\,=\,0.421$ for 
$I(\mu)/I(1)\,=\,1\,-\,$a$\left(1\,-\,\mu\right)\,-\,b\left(1\,-\,\mu\right)^{2}$, with $\mu$ being the cosine of the angle between the line of sight and the surface normal. For the secondary 
star we used the Tables of \citet{claret+bloemen11-1}. We interpolated between the values of V and R for a $T\,=\,3000\,\mathrm{K}$ and $\mathrm{log}\,g\,=\,5$ star, to obtain quadratic limb 
darkening coefficients a$'\,=\,0.62$ and $b'\,=\,0.273$. All other parameters were allowed to vary.

\subsection{Minimisation}

Initial minimisation is achieved using the downhill-{\sc simplex} and {\sc levenberg-marquardt} methods \citep{press02-1}, while the Markov Chain Monte Carlo (MCMC) method \citep{pressetal07-1}
was used to determine the distributions of our model parameters \citep[e.g.][and references therein]{ford06-1}.

The MCMC method involves making random jumps in the model parameters, with new models being accepted or rejected according to their probability computed as a Bayesian posterior probability
(the probability of the model parameters, $\theta,$ given the data, D, $P\left(\theta|D\right)$). $P\left(\theta|D\right)$ is driven by a combination of $\chi^2$ and a prior
probability, $P(\theta)$, that is based on previous knowledge of the model parameters. 

In our case, the prior probabilities for most parameters are assumed to be uniform. The photometric data provide constraints for the radii and inclination angle, however, the
photometry alone cannot constrain the masses, as the light curve itself is only weakly depended on $q$. To alleviate this, we can use our knowledge of \kwd\, and \ksec. At each jump,
the model values \kwdm\, and \ksecm\, are calculated through $q$, $i$ and \vscale. $P(\theta)$ is then evaluated on the basis of the observed \kwd\, and \ksec, assuming a Gaussian prior 
probability $P(\mu,\sigma^{2})$, with $\mu$ and $\sigma$ corresponding to the measured values and errors of \kwd\, and \ksec.

A crucial practical consideration of MCMC is the number of steps required to fairly sample the parameter space, which is largely determined by how closely the distribution of parameter jumps 
matches the true distribution. We therefore built up an estimate of the correct distribution starting from uncorrelated jumps in the parameters, after which we computed the covariance matrix 
from the resultant chain of parameter values. The covariance matrix was then used to define a multivariate normal distribution that was used to make the jumps for the next chain. At each stage 
the actual size of the jumps was scaled by a single factor set to deliver a model acceptance rate of $\approx 25$\% \citep{robertsetal97-1}. After 3 such cycles, the covariance matrix showed 
only small changes, and at this point we carried out the long "production runs" during which the covariance and scale factor which define the parameter jumps were held fixed.

\subsection{Stellar parameters}

Using the following set of equations, the stellar and binary parameters are obtained directly from the posterior distribution of the model parameters, as outputed from the MCMC minimisation. 

\noindent
The binary separation is obtained from the model parameter \vscale\, through

\begin{equation}
 \label{eq:binsep}
a\,=\,\frac{\porb}{2\,\pi}\,\vscale
\end{equation}

\noindent
The white dwarf and secondary masses are obtained from the model parameters $q$ and \vscale\, as

\begin{equation}
 \label{eq:wdmass}
\mwd\,=\,\frac{\porb}{2\,\pi\,\mathrm{G}}\,\frac{1}{1\,+\,q}\,\vscale^{3}
\end{equation}

\noindent
and

\begin{equation}
 \label{eq:secmass}
\msec\,=\,\frac{\porb}{2\,\pi\,\mathrm{G}}\,\frac{q}{1\,+\,q}\,\vscale^{3}
\end{equation}

\noindent
The stellar radii are directly obtained from the model parameters \rwda\, and \rseca\, and Eq.\,\ref{eq:binsep} and the surface gravity of the white dwarf is of course given by 

\begin{equation}
 \label{eq:logg}
\mathrm{log}\,g\,=\,\rm{log}\left(\frac{G\mwd}{R^{2}_{\rm{WD}}}\right)
\end{equation}

\subsection{Intrinsic data uncertainties}

The acquisition of high-precision absolute photometry on the LT in service mode is somewhat difficult to achieve. Each observing block individually covered only a third of the orbital phase 
and the blocks were obtained over many nights, under varying conditions (seeing, sky brightness, extinction, airmass). The data are sensitive to changes in conditions, as they have been 
obtained through the very broad and non-standard V+R filter of RISE. In the absence of a flux standard, the photometry cannot be calibrated in absolute terms. When phase-folding the LT data, 
significant scatter is found at orbital phases where individual observing blocks with discrepant calibrations contribute. This affects both the shape of the eclipse, mainly the steepness of 
the WD ingress/egress and, to a lesser extend, the eclipse duration, and the out-of-eclipse variation, i.e. the profile of the ellipsoidal modulation. As a result, there is an unavoidable 
systematic uncertainty in the photometric accuracy of our data, which will influence the determination of the stellar parameters.

To gauge the effect of the systematic uncertainties we worked in the following fashion: each observing block has been reduced thrice, each time using one of the three comparison stars reported 
in Table\,\ref{tab:coordmag}. C1 has a $g-r$ colour index comparable to \system, C2 is fairly red, while C3 is fairly blue. The data of each reduction were then phase-folded together and two 
light curves were produced: one containing all the photometric points and one where (2-3) observing blocks with an obviously large intrinsic scattering were omitted. Thus, we ended up with six 
phase-folded light curves. A dedicated MCMC optimisation was calculated for each light curve. We will use the following notation when refering to these chains: C1A denotes a light curve 
produced with comparison star C1 and all data points, C2E denotes a light curve produced with comparison star C2 excluding observing blocks, and so on.


\section{Results}
\label{sec:result}

\begin{figure*}
\centering
 \includegraphics[width=\textwidth]{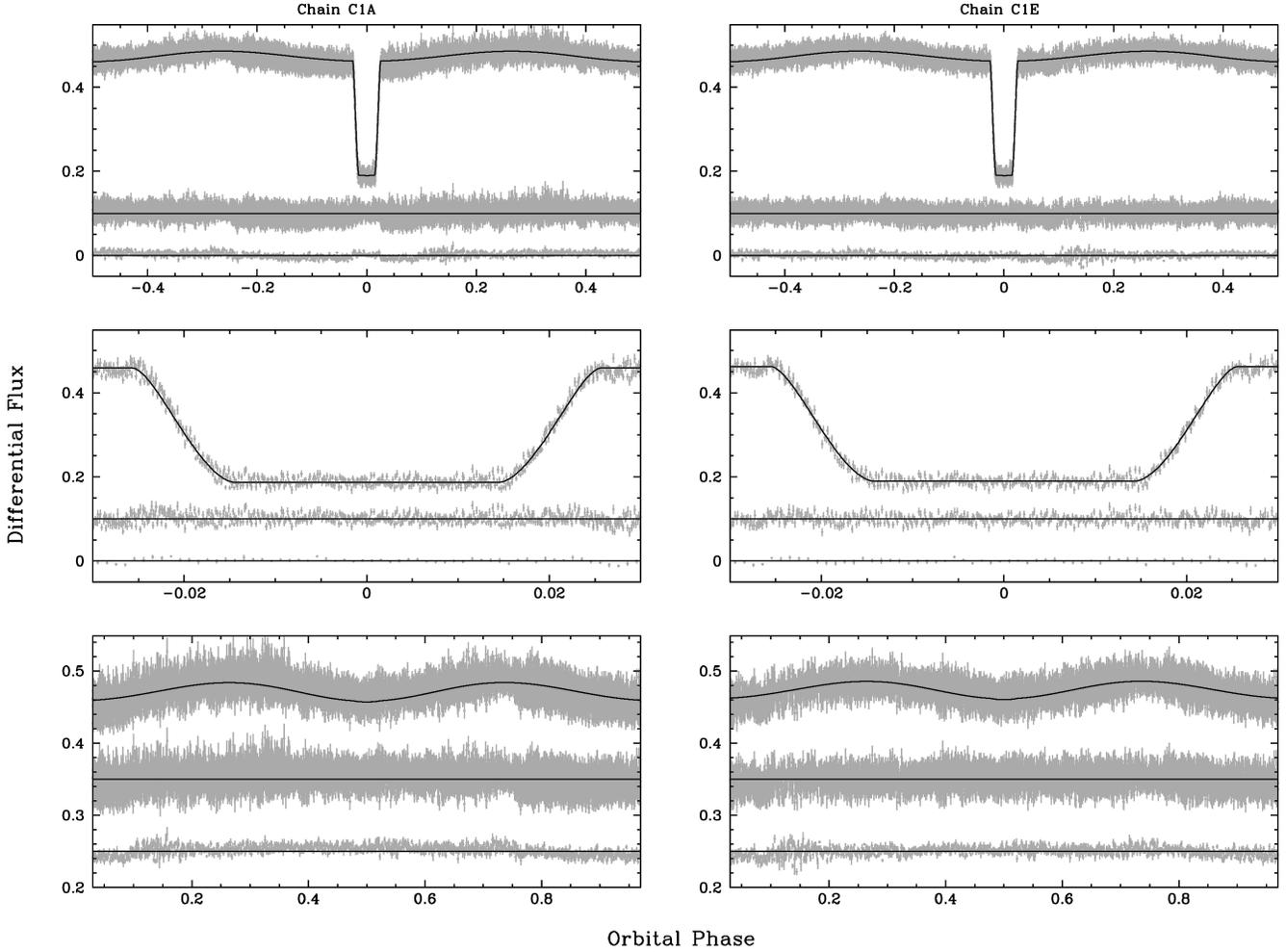}
  \caption{Light curve fitting results for models C1A (left) and C1E (right). In each of the six panels we plot the phase-folded light curve with the model superimposed (top trace), the 
residuals of the fit (middle trace, offseted from 0 for clarity) and a binned version of the residuals (bottom trace). Shown are the entire light curve (top panels), a zoom around the 
eclipse (middle panels) and the out-of-eclipse ellipsoidal modulation (bottom panels).}
  \label{fig:fitfig}
\end{figure*}

The results of the six MCMC processes are summarised in Table\,\ref{tab:mcmcresult}. The quoted values and errors are purely of statistical nature and represent the mean and RMS of the 
posterior distribution of each parameter. The radius of the secondary, as determined by \rseca\, and $a$, is measured along the line connecting the centres of the two stars and, due to 
the tidal distortion, its value is larger than the average radius. Therefore, on Table\,\ref{tab:mcmcresult} we also report the more representative value of the volume-averaged radius.

\begin{table*}
\setlength{\tabcolsep}{0.8ex}
\centering
\caption{Stellar and binary parameters obtained from MCMC optimisation. The quoted values and errors are the mean and RMS of the posterior distribution of each parameter. The chains represent
light curves created using comparison stars C1, C2 or C3 and either including all (A) observing blocks or excluding (E) those with obviously large scattering. See text for details.}
\label{tab:mcmcresult}
\begin{tabular}{@{}ccccccc@{}}
\hline
Parameter & C1A & C1E & C2A & C2E & C3A & C3E \\
\hline
q & $0.380\,\pm\,0.010$ & $0.380\,\pm\,0.010$ & $0.381\,\pm\,0.010$ & $0.380\,\pm\,0.010$ & $0.378\,\pm\,0.010$ & $0.379\,\pm\,0.010$ \\
i [$^{\circ}$] & $79.05\,\pm\,0.15$ & $79.28\,\pm\,0.15$ & $79.03\,\pm\,0.15$ & $79.13\,\pm\,0.15$ & $79.36\,\pm\,0.18$ & $79.29\,\pm\,0.16$ \\
\mwd\, [\Msun] & $0.415\,\pm\,0.010$ & $0.414\,\pm\,0.010$ & $0.415\,\pm\,0.010$ & $0.415\,\pm\,0.010$ & $0.414\,\pm\,0.010$ & $0.414\,\pm\,0.010$ \\
\rwds\, [\Rsun] & $0.0157\,\pm\,0.0003$ & $0.0159\,\pm\,0.0003$ & $0.0161\,\pm\,0.0003$ & $0.0159\,\pm\,0.0003$ & $0.0138\,\pm\,0.0003$ & $0.0150\,\pm\,0.0003$ \\
WD $\mathrm{log}\,g$ & $7.664\,\pm\,0.015$ & $7.652\,\pm\,0.016$ & $7.641\,\pm\,0.015$ & $7.649\,\pm\,0.017$ & $7.773\,\pm\,0.023$ & $7.700\,\pm\,0.019$ \\
\msec\, [\Msun] & $0.158\,\pm\,0.006$ & $0.157\,\pm\,0.006$ & $0.158\,\pm\,0.006$ & $0.158\,\pm\,0.006$ & $0.156\,\pm\,0.007$ & $0.157\,\pm\,0.006$ \\
\rsecs\, [\Rsun] & $0.217\,\pm\,0.003$ & $0.212\,\pm\,0.003$ & $0.217\,\pm\,0.003$ & $0.215\,\pm\,0.003$ & $0.210\,\pm\,0.004$ & $0.211\,\pm\,0.003$ \\
\rsecv\, [\Rsun] & $0.202\,\pm\,0.003$ & $0.199\,\pm\,0.003$ & $0.203\,\pm\,0.003$ & $0.201\,\pm\,0.003$ & $0.197\,\pm\,0.003$ & $0.198\,\pm\,0.003$\\
\tsec\, [$\rm{K}$] & $\sim\,2530$ & $\sim\,2550$ & $\sim\,2530$ & $\sim\,2550$ & $\sim\,2500$ & $\sim\,2550$ \\
Binary separation [\Rsun] & $0.871\,\pm\,0.008$ & $0.870\,\pm\,0.008$ & $0.871\,\pm\,0.008$ & $0.871\,\pm\,0.008$ & $0.869\,\pm\,0.008$ & $0.870\,\pm\,0.008$ \\
\hline
\end{tabular}
\end{table*}

To illustrate the achieved quality of the fits, we plot models C1A and C1E in Figure\,\ref{fig:fitfig}. While the overall quality of the fit is very satisfactory, the model seems to slightly
overpredict the flux at the ``wings'' of the ellipsoidal modulation profile (phases $\sim\,0.05\,-\,0.15$ and $\sim\,0.85\,-\,0.95$). This discrepancy could be data related, due to the 
intrinsic scattering of points; system related, e.g. due to the presence of starspots affecting the modulation; model related, as the treatment of stellar temperatures is based on blackbody 
spectra, for one specific wavelength; or due to a combination of these factors.

With regard to the binary and stellar parameters, the MCMC results indicate the following: as expected for a detached system, the light curves depend very weakly on $q$ and its value is well
constrained by the radial velocitites. All six chains give inclination angle values just above $79^{\circ}$, consistent with each other within the errors. There is a slight shift upwards when
excluding blocks from the phase-folded light curve.

The tight spectroscopic constraints, mean that the component masses are largely independent of the model/data set used. Thus, the white dwarf in 
\system\, has a mass of $\mwd\,=\,0.415\,\pm\,0.010\,\Msun$ and the secondary star a mass of $\msec\,=\,0.158\,\pm\,0.006\,\Msun$. 

\begin{figure}
\centering
 \includegraphics[angle=-90,width=84mm]{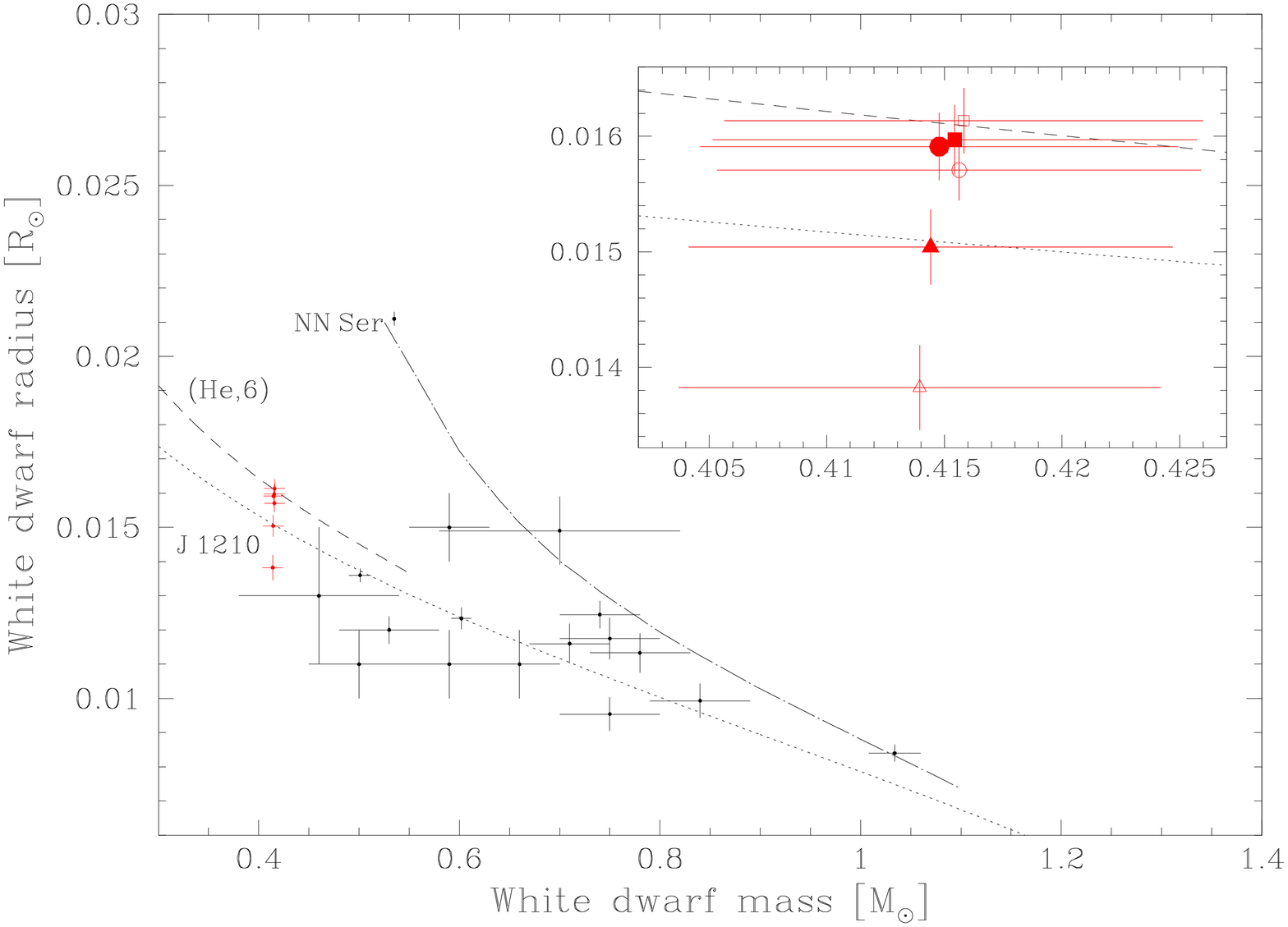}
  \caption{Mass-radius plot for white dwarfs. Black points are data from \citet{provencaletal98-1}, \citet{provencaletal02-1} and \citet{casewelletal09-1}. The dotted line is the 
zero-temperature mass-radius relation of Eggleton as quoted in \citet{verbunt+rappaport88-1}. The dashed line, marked as (He,6) is a M-R relation for a $\twd\,=\,6000\,\rm{K}$, He-core WD, 
with a hydrogen layer of $M(\rm{H})/\mwd\,=\,3\times10^{-4}$, interpolated from the models of \citet{althaus+benvenuto97-1}. NN\,Ser \citep{parsonsetal10-1} is marked, along with the track
for $\twd\,=\,60000\,\rm{K}$, C/O-core WD, $M(\rm{H})/\mwd\,=\,10^{-4}$ (long dash-dot line), indicating the accuracy achieved in eclipsing PCEBs. The results of the six chains for \system\, 
are plotted in red (online version only). Inset panel: zoom-in on the values of \system. The points are C1A: open circle; C1E: filled circle; C2A: open square; C2E: filled square; C3A: open 
triangle; C3E: filled triangle.}
  \label{fig:wdmr}
\end{figure}

The quantity most seriously affected by systematics is the white dwarf radius. This is especially evident when considering models C3A and C3E. However, such a discrepancy is expected, since 
C3 is considerably bluer than \system\, and is more susceptible to airmass/colour effects, leading to large intrinsic scattering. The values for \rwds\, as obtained from C1A, C1E, C2A and C2E 
are consistent within their errors, indicating a systematic uncertainty comparable to the statistical one. This is illustrated in Figure\,\ref{fig:wdmr}.

\begin{figure}
\centering
 \includegraphics[angle=-90,width=84mm]{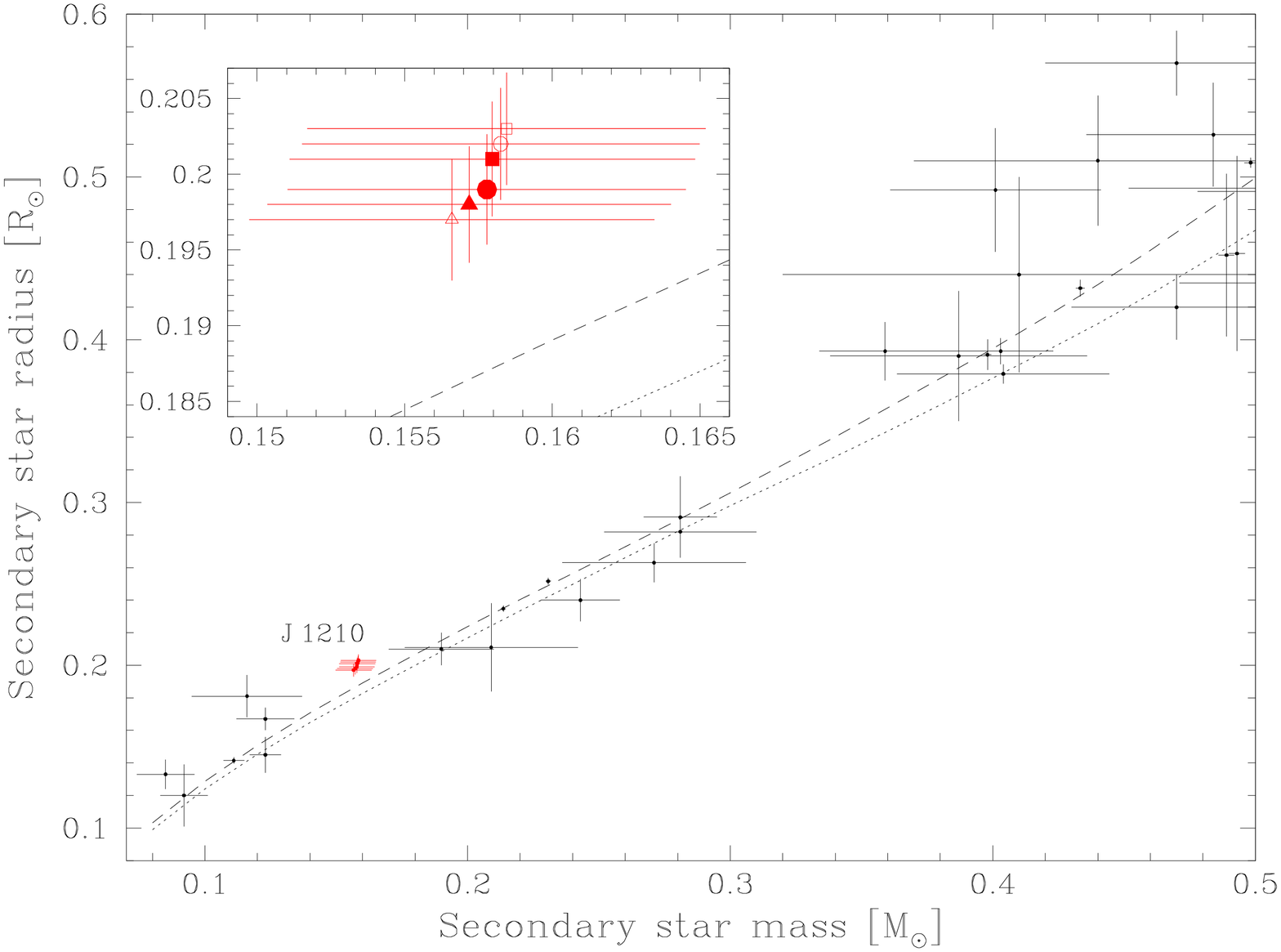}
  \caption{Mass-radius plot for low-mass stars. Black points are data from \citet{lopez-morales07-1} and \citet{beattyetal07-1}, where the masses of single stars were determined using 
mass-luminosity relations. The dotted line is the 5.0-Gyr isochrone from \citet{baraffeetal98-1}. The dashed line is a 5.0-Gyr model including effects of magnetic activity from 
\citet{moralesetal10-1}. The results of the six chains for the volume-averaged radius of the secondary in \system\, are plotted in red (online version only). Inset panel: zoom-in on the values 
of \system. The points are C1A: open circle; C1E: filled circle; C2A: open square; C2E: filled square; C3A: open triangle; C3E: filled triangle.}
  \label{fig:secmr}
\end{figure}

The secondary star radius is affected in a similar, albeit less pronounced, way. All six models lead to values broadly consistent within their statistical errors and a systematic uncertainty
of the same order as the statistical one. Figure\,\ref{fig:secmr} shows the six different values of the volume-averaged secondary star radius overplotted on a M-R relation for MS stars. Taken 
at face value, the results of the MCMC optimisation indicate that the secondary is $\sim\,10$ percent larger than theoretically predicted. As can be seen in Figure\,\ref{fig:secmr}, this
discrepancy drops to $\sim\,5$ percent, if magnetic activity of the secondary is taken into account. With regard to the secondary temperature, we note again that due to the blackbody 
approximation, the value of \tsec\, does not necessarily represent the true temperature of the star, it is effectively just a flux scaling factor.

The gravitational redshift predicted by the light curve models (Table\,\ref{tab:mcmcresult}), correcting for the redshift of the secondary star, the difference in transverse Doppler shifts and 
the potential at the secondary star owing to the white dwarf, are $\zphot\,=\,15.9\,\pm\,0.4\,\kms$ from C1E and $\zphot\,=\,15.8\,\pm\,0.4\,\kms$ from C2E, where the errors are purely 
statistical and have been derived in the same manner as the other quantities reported in Table\,\ref{tab:mcmcresult}. The systematic uncertainties in our photometric data might still be 
influencing the result, as the inclination angle and the stellar radii enter the calculation of \zphot. Comparing \zphot\, with the spectroscopically determined value of 
$\zspec\,=\,\gamwd\,-\,\gamsec\,=\,11.9\,\pm\,1.7\,\kms$ we find that they are consistent within $\sim\,2\,\sigma$. The systemic velocities \gamwd\, and \gamsec\, are determined from 
spectroscopic observations obtained using a dual-arm spectrograph, with the white dwarf velocity measured in the blue arm and that of the secondary measured in the red arm 
(Sec.\,\ref{sec:obsnred}, \ref{sec:radvel}). The observations in both arms are independently wavelength-calibrated and the RMS of $\sim\,$0.03\AA\, (Sec.\,\ref{sec:obsnred}) corresponds to an 
accuracy of the zero-point of $\sim\,1\,-\,2\,\kms$. The potential of an offset in the calibrations of the two arms enters the determination of \zspec\, as an additional systematic uncertainty.


\section{Past and future evolution of \system}
\label{sec:evolution}

Considering its short orbital period, \system\, must have formed through common-envelope evolution \citep{paczynski76-1,webbink08-1}. As shown by \citet{schreiber+gaensicke03-1}, if the binary
and stellar parameters are known, it is possible to reconstruct the past and predict the future evolution of PCEBs for a given angular momentum loss prescription. Here, we assume classical
disrupted magnetic braking \citep{verbunt+zwaan81-1}. In this context, given the low mass of the secondary, the only angular momentum loss mechanism for \system\, is gravitational 
radiation. Based on the temperature and the mass of the white dwarf we interpolate the cooling tracks of \citet{althaus+benvenuto97-1} and obtain a cooling age of 
$t_{\rm{cool}}\,=\,3.5\,\rm{Gyr}$. This corresponds to the time that passed since the binary left the common envelope. We calculate the period it had when it left the common envelope to be 
$P_{\mathrm{CE}}\,=\,4.24$h. Following the same method as in \citet{zorotovicetal10-1} and based on their results we reconstructed the initial parameters of the binary using a common-envelope 
efficiency of $\alpha_{\mathrm{CE}}=0.25$ and the same fraction of recombination energy \citep[see][for more details]{zorotovicetal10-1}. We found an initial mass of $M_{\rm{prog}}\,=\,1.33\,\Msun$ 
for the progenitor of the white dwarf, which filled its Roche lobe when its radius was $R_{\rm{prog}}\,=\,91.3\,\Rsun$. At that point, the orbital separation was $a\,=\,162.7\,\Rsun$, and the 
age of the system was $t_{\rm{sys}}\,=\,4.4\,\rm{Gyr}$, since the time it was formed. Using the radius of the secondary\footnote{We assume a representative value of $\rsecv\,=\,0.2\,\Rsun$ 
for the volume-averaged radius of the secondary} we calculate that the system will reach a semi-detached configuration and become a cataclysmic variable (CV) at an orbital period of 
$P_{\rm{sd}}\,\sim\,2\,$h in $t_{\rm{sd}}\,=\,1.5\,\rm{Gyr}$.

Given that the current \porb\, places \system\, right at the upper edge of the CV orbital period gap\footnote{The orbital period range where only a small number of CVs are found.}, 
and that the calculated $P_{\rm{sd}}$, when \system\, will start mass-transfer, is right 
at the lower edge of the period gap, we are tempted to speculate whether \system\, is in fact a detached CV entering (or just having entered) the period gap. \citet{davisetal08-1} have shown 
that a large number of detached WD+MS binaries with orbital periods between 2-3 hours are in fact CVs that have switched off mass-transfer and are crossing the period gap. This could in 
principle explain the apparently over-sized secondary in \system, as expected from the disrupted magnetic braking theory \citep[e.g.][]{rappaportetal83-1}. However, the temperature of the WD
in \system\, seems to be uncomfortably low for a WD that has recently stopped accreting \citep{townsley+gaensicke09-1}.


\section{Discussion and Conclusions}
\label{sec:conclusion}

In this paper, we have identified \system\, as an eclipsing PCEB containing a very cool, low-mass, DAZ white dwarf and a low-mass main-sequence companion. 

Using combined constraints from spectroscopic and photometric observations we have managed to measure the fundamental stellar parameters of the binary components. Systematic uncertainties in 
the absolute calibration of our photometric data, influence the determination of the stellar radii. The stellar masses, however, remain unaffected and were measured to a $1\%$ accuracy. The
(formal) statistical uncertainties in all binary parameters indicate the level of precision that can be achieved in this system. All parameters are summarised in Table\,\ref{tab:allparam}.

\begin{table}
\setlength{\tabcolsep}{0.7ex}
\centering
\caption{Adopted stellar and binary parameters for \system.}
\label{tab:allparam}
\begin{tabular}{@{}cc@{}}
\hline 
Parameter & Value \\
\hline 
\porb\,[d] &  0.124\,489\,764(1) \\
$q$ & $0.379\,\pm\,0.009$ \\
$a$\,[\Rsun] & $0.870\,\pm\,0.008$ \\
Inclination [$^{\circ}$] & $(79.05\,-\,79.36)\,\pm\,0.15$\\
\mwd\,[\Msun] & $0.415\,\pm\,0.010$  \\
\rwds\,[\Rsun] & $(0.0157\,-\,0.0161)\,\pm\,0.0003$ \\ 
$\mathrm{log}\,g$ & $7.65\,\pm\,0.02$ \\
\twd\,[$\rm{K}$] & $6000\,\pm\,200$ \\
\kwd\,[\kms] & $95.3\,\pm\,2.1$ \\
\msec\,[\Msun] & $0.158\,\pm\,0.006$ \\ 
\rsecs\,[\Rsun] & $(0.210\,-\,0.217)\,\pm\,0.003$ \\
\rsecv\,[\Rsun] & $(0.197\,-\,0.203)\,\pm\,0.003$ \\
\ksec\,[\kms] & $251.7\,\pm\,2.0$ \\
\hline
\end{tabular}
\end{table}

With a mass of $\mwd\,=\,0.415\,\pm\,0.010\,\Msun$ and a temperature of $\twd\,\sim\,6000\,\mathrm{K}$, the DAZ white dwarf in \system\, pushes the boundaries in a hitherto unexplored region of 
the WD parameter space. The M-R results from the four Chains C1 and C2 are consistent with a He-core WD, assuming a hydrogen layer of $M(\rm{H})/\mwd\,=\,3\times10^{-4}$. However, due to lack 
of observational constraints for the H-layer thickness and the uncertainty in the radii, we will defer identifying the WD as a definite He-core and simply emphasise the strong candidacy. 

The secondary star, with a mass of $\msec\,=\,0.158\,\pm\,0.006\,\Msun$, illustrates once more the excellent opportunity that PCEBs give us for testing and calibrating the M-R 
relations of low-mass stars. Taking the radius measurements at face value, the secondary star seems to be $\sim\,10$ percent larger than the theoretical values, although this drops to 
$\sim\,5$ percent, if magnetic activity is taken into consideration. In this context, the magnetic activity present in the secondary can lead to the formation 
of stellar (dark) spots on the surface. The effect of these spots is to block the outgoing heat flux, reducing $T_{\rm{eff}}$ and, as a result, the secondary expands to maintain 
thermal equilibrium \citep{chabrieretal07-1,moralesetal10-1}. \citet{krausetal11-1} found that low-mass stars in short period binaries appear to be overinflated (although their 
analysis was restricted to $\msec\,>\,0.3\,\Msun$), which seems to be the case for \system. We should note however, that the mass and radius of the secondary star in the eclipsing PCEB NN\,Ser 
(with $\msec\,=\,0.111\,\pm\,0.004\,\Msun$ and comparable orbital period to \system) is consistent with theoretical M-R predictions, even though it is heavily irradiated by the hot WD primary 
\citep{parsonsetal10-1}. 

We have speculated whether \system\, is in fact a detached CV entering the period gap, which could explain the large radius of the secondary. This hypothesis could be tested by measuring the 
rotational velocity of the white dwarf. This can be achieved through high-resolution spectroscopy of the \Ion{Fe}{I} absorption lines in the WD photosphere \citep[see e.g.][]{tappertetal11-2}.

In any case, it is highly desirable to  improve the measurement of the stellar radii in SDSS1210 to the comparable precision to the masses presented here. This will require high-precision 
photometry in standard filters, such as e.g. delivered by ULTRACAM \citep{dhillonetal07-1}.


\section*{Acknowledgements}
We thank the anonymous referee for a prompt report. BTG, TRM, EB and CMC are supported by an STFC Rolling Grant. MRS and ARM acknowledge financial support from FONDECYT in the form of grants 
1100782 and 3110049. MZ acknowledges support from Gemini/CONICYT (grant 32100026). Based in part on observations made with the William Herschel Telescope operated on the island of La Palma by 
the Isaac Newton Group in the Spanish Observatorio del Roque de los Muchachos of the Instituto de Astrof\'isica de Canarias and on observations made with the Liverpool Telescope operated on 
the island of La Palma by Liverpool John Moores University in the Spanish Observatorio del Roque de los Muchachos of the Instituto de Astrofisica de Canarias with financial support from the 
UK Science and Technology Facilities Council.


\bibliographystyle{mn_new} 
\bibliography{aamnem99,thesis}

\begin{thebibliography}{67}
\expandafter\ifx\csname natexlab\endcsname\relax\def\natexlab#1{#1}\fi

\bibitem[{{Abazajian} et~al.(2009)}]{abazajianetal09-1}
{Abazajian}, K.~N., et~al., 2009, ApJS, 182, 543

\bibitem[{{Adelman-McCarthy} et~al.(2008)}]{adelmanetal08-1}
{Adelman-McCarthy}, J.~K., et~al., 2008, ApJS, 175, 297

\bibitem[{{Althaus} \& {Benvenuto}(1997)}]{althaus+benvenuto97-1}
{Althaus}, L.~G., {Benvenuto}, O.~G., 1997, ApJ, 477, 313

\bibitem[{{Andersen}(1991)}]{andersen91-1}
{Andersen}, J., 1991, ARA\&A, 3, 91

\bibitem[{{Baraffe} et~al.(1998){Baraffe}, {Chabrier}, {Allard}, \&
  {Hauschildt}}]{baraffeetal98-1}
{Baraffe}, I., {Chabrier}, G., {Allard}, F., {Hauschildt}, P.~H., 1998, A\&A,
  337, 403

\bibitem[{{Bayless} \& {Orosz}(2006)}]{bayless+orosz06-1}
{Bayless}, A.~J., {Orosz}, J.~A., 2006, ApJ, 651, 1155

\bibitem[{{Beatty} et~al.(2007)}]{beattyetal07-1}
{Beatty}, T.~G., et~al., 2007, ApJ, 663, 573

\bibitem[{{Berger} et~al.(2006)}]{bergeretal06-1}
{Berger}, D.~H., et~al., 2006, ApJ, 644, 475

\bibitem[{{Bertin} \& {Arnouts}(1996)}]{bertin+arnouts96-1}
{Bertin}, E., {Arnouts}, S., 1996, A\&AS, 117, 393

\bibitem[{{Brown} et~al.(2011){Brown}, {Kilic}, {Hermes}, {Allende Prieto},
  {Kenyon}, \& {Winget}}]{brownetal11-1}
{Brown}, W.~R., {Kilic}, M., {Hermes}, J.~J., {Allende Prieto}, C., {Kenyon},
  S.~J., {Winget}, D.~E., 2011, ArXiv eprints., 1107.2389

\bibitem[{{{\c C}ak{\i}rl{\i}} \& {Ibano{\v g}lu}(2010)}]{cakirli+ibanoglu10-1}
{{\c C}ak{\i}rl{\i}}, {\"O}., {Ibano{\v g}lu}, C., 2010, MNRAS, 401, 1141

\bibitem[{{Casewell} et~al.(2009){Casewell}, {Dobbie}, {Napiwotzki},
  {Burleigh}, {Barstow}, \& {Jameson}}]{casewelletal09-1}
{Casewell}, S.~L., {Dobbie}, P.~D., {Napiwotzki}, R., {Burleigh}, M.~R.,
  {Barstow}, M.~A., {Jameson}, R.~F., 2009, MNRAS, 395, 1795

\bibitem[{{Chabrier} et~al.(2007){Chabrier}, {Gallardo}, \&
  {Baraffe}}]{chabrieretal07-1}
{Chabrier}, G., {Gallardo}, J., {Baraffe}, I., 2007, A\&A, 472, L17

\bibitem[{{Claret} \& {Bloemen}(2011)}]{claret+bloemen11-1}
{Claret}, A., {Bloemen}, S., 2011, A\&A, 529, A75+

\bibitem[{{Copperwheat} et~al.(2010){Copperwheat}, {Marsh}, {Dhillon},
  {Littlefair}, {Hickman}, {G{\"a}nsicke}, \&
  {Southworth}}]{copperwheatetal10-1}
{Copperwheat}, C.~M., {Marsh}, T.~R., {Dhillon}, V.~S., {Littlefair}, S.~P.,
  {Hickman}, R., {G{\"a}nsicke}, B.~T., {Southworth}, J., 2010, MNRAS, 402,
  1824

\bibitem[{{Davis} et~al.(2008){Davis}, {Kolb}, {Willems}, \&
  {G{\"a}nsicke}}]{davisetal08-1}
{Davis}, P.~J., {Kolb}, U., {Willems}, B., {G{\"a}nsicke}, B.~T., 2008, MNRAS,
  389, 1563

\bibitem[{{Debes}(2006)}]{debes06-1}
{Debes}, J.~H., 2006, ApJ, 652, 636

\bibitem[{{Dhillon} et~al.(2007)}]{dhillonetal07-1}
{Dhillon}, V.~S., et~al., 2007, MNRAS, 378, 825

\bibitem[{{Dimitrov} \& {Kjurkchieva}(2010)}]{dimitrov10-1}
{Dimitrov}, D.~P., {Kjurkchieva}, D.~P., 2010, MNRAS, 406, 2559

\bibitem[{{Drake} et~al.(2010)}]{drakeetal10-1}
{Drake}, A.~J., et~al., 2010, ArXiv eprints., 1009.3048

\bibitem[{{Dupuis} et~al.(1993){Dupuis}, {Fontaine}, {Pelletier}, \&
  {Wesemael}}]{dupuisetal93-1}
{Dupuis}, J., {Fontaine}, G., {Pelletier}, C., {Wesemael}, F., 1993, ApJS, 84,
  73

\bibitem[{{Ford}(2006)}]{ford06-1}
{Ford}, E.~B., 2006, ApJ, 642, 505

\bibitem[{{G\"ansicke} et~al.(2004){G\"ansicke}, {Araujo-Betancor}, {Hagen},
  {Harlaftis}, {Kitsionas}, {Dreizler}, \& {Engels}}]{gaensickeetal04-1}
{G\"ansicke}, B.~T., {Araujo-Betancor}, S., {Hagen}, H.-J., {Harlaftis}, E.~T.,
  {Kitsionas}, S., {Dreizler}, S., {Engels}, D., 2004, A\&A, 418, 265

\bibitem[{{Horne}(1986)}]{horne86-1}
{Horne}, K., 1986, PASP, 98, 609

\bibitem[{{Irwin} et~al.(2010)}]{irwinetal10-1}
{Irwin}, J., et~al., 2010, ApJ, 718, 1353

\bibitem[{{Koester}(2010)}]{koester10-1}
{Koester}, D., 2010, MmSAI, 81, 921

\bibitem[{{Koester} \& {Wilken}(2006)}]{koester+wilken06-1}
{Koester}, D., {Wilken}, D., 2006, A\&A, 453, 1051

\bibitem[{{Kraus} et~al.(2011){Kraus}, {Tucker}, {Thompson}, {Craine}, \&
  {Hillenbrand}}]{krausetal11-1}
{Kraus}, A.~L., {Tucker}, R.~A., {Thompson}, M.~I., {Craine}, E.~R.,
  {Hillenbrand}, L.~A., 2011, ApJ, 728, 48

\bibitem[{{L{\'o}pez-Morales}(2007)}]{lopez-morales07-1}
{L{\'o}pez-Morales}, M., 2007, ApJ, 660, 732

\bibitem[{{Marsh}(1989)}]{marsh89-1}
{Marsh}, T.~R., 1989, PASP, 101, 1032

\bibitem[{{Morales} et~al.(2008){Morales}, {Ribas}, \&
  {Jordi}}]{moralesetal08-1}
{Morales}, J.~C., {Ribas}, I., {Jordi}, C., 2008, A\&A, 478, 507

\bibitem[{{Morales} et~al.(2010){Morales}, {Gallardo}, {Ribas}, {Jordi},
  {Baraffe}, \& {Chabrier}}]{moralesetal10-1}
{Morales}, J.~C., {Gallardo}, J., {Ribas}, I., {Jordi}, C., {Baraffe}, I.,
  {Chabrier}, G., 2010, ApJ, 718, 502

\bibitem[{{Morales} et~al.(2009)}]{moralesetal09-1}
{Morales}, J.~C., et~al., 2009, ApJ, 691, 1400

\bibitem[{{Morrissey} et~al.(2007)}]{morrisseyetal07-1}
{Morrissey}, P., et~al., 2007, 173, 682

\bibitem[{{Nebot G{\'o}mez-Mor{\'a}n} et~al.(2009)}]{nebotgomezmoranetal09-1}
{Nebot G{\'o}mez-Mor{\'a}n}, A., et~al., 2009, A\&A, 495, 561

\bibitem[{{Paczynski}(1976)}]{paczynski76-1}
{Paczynski}, B., 1976, in {P.~Eggleton, S.~Mitton, \& J.~Whelan}, ed.,
  Structure and Evolution of Close Binary Systems, vol.~73 of \emph{IAU
  Symposium}, p.~75

\bibitem[{{Panei} et~al.(2000){Panei}, {Althaus}, \&
  {Benvenuto}}]{paneietal00-2}
{Panei}, J.~A., {Althaus}, L.~G., {Benvenuto}, O.~G., 2000, A\&A, 353, 970

\bibitem[{{Parsons} et~al.(2010{\natexlab{a}}){Parsons}, {Marsh},
  {Copperwheat}, {Dhillon}, {Littlefair}, {G{\"a}nsicke}, \&
  {Hickman}}]{parsonsetal10-1}
{Parsons}, S.~G., {Marsh}, T.~R., {Copperwheat}, C.~M., {Dhillon}, V.~S.,
  {Littlefair}, S.~P., {G{\"a}nsicke}, B.~T., {Hickman}, R.,
  2010{\natexlab{a}}, MNRAS, 402, 2591

\bibitem[{{Parsons} et~al.(2011){Parsons}, {Marsh}, {G{\"a}nsicke}, {Drake}, \&
  {Koester}}]{parsonsetal11-1}
{Parsons}, S.~G., {Marsh}, T.~R., {G{\"a}nsicke}, B.~T., {Drake}, A.~J.,
  {Koester}, D., 2011, ApJ Lett., 735, L30+

\bibitem[{{Parsons} et~al.(2010{\natexlab{b}})}]{parsonsetal10-2}
{Parsons}, S.~G., et~al., 2010{\natexlab{b}}, MNRAS, 407, 2362

\bibitem[{{Press}(2002)}]{press02-1}
{Press}, W.~H., 2002, {Numerical recipes in C++ : the art of scientific
  computing}, Cambridge Univ. Press, Cambridge

\bibitem[{{Press} et~al.(2007){Press}, {Teukolsky}, {Vetterling}, \&
  {Flannery}}]{pressetal07-1}
{Press}, W.~H., {Teukolsky}, A.~A., {Vetterling}, W.~T., {Flannery}, B.~P.,
  2007, {Numerical recipes. The art of scientific computing, 3rd edn.},
  Cambridge: University Press

\bibitem[{{Provencal} et~al.(1998){Provencal}, {Shipman}, {Hog}, \&
  {Thejll}}]{provencaletal98-1}
{Provencal}, J.~L., {Shipman}, H.~L., {Hog}, E., {Thejll}, P., 1998, ApJ, 494,
  759

\bibitem[{{Provencal} et~al.(2002){Provencal}, {Shipman}, {Koester},
  {Wesemael}, \& {Bergeron}}]{provencaletal02-1}
{Provencal}, J.~L., {Shipman}, H.~L., {Koester}, D., {Wesemael}, F.,
  {Bergeron}, P., 2002, ApJ, 568, 324

\bibitem[{{Pyrzas} et~al.(2009)}]{pyrzasetal09-1}
{Pyrzas}, S., et~al., 2009, MNRAS, 394, 978

\bibitem[{{Rappaport} et~al.(1983){Rappaport}, {Verbunt}, \&
  {Joss}}]{rappaportetal83-1}
{Rappaport}, S., {Verbunt}, F., {Joss}, P.~C., 1983, ApJ, 275, 713

\bibitem[{{Rebassa-Mansergas} et~al.(2010){Rebassa-Mansergas}, {G{\"a}nsicke},
  {Schreiber}, {Koester}, \& {Rodr{\'{\i}}guez-Gil}}]{rebassamansergasetal10-1}
{Rebassa-Mansergas}, A., {G{\"a}nsicke}, B.~T., {Schreiber}, M.~R., {Koester},
  D., {Rodr{\'{\i}}guez-Gil}, P., 2010, MNRAS, 402, 620

\bibitem[{{Rebassa-Mansergas} et~al.(2011){Rebassa-Mansergas}, {Nebot
  G{\'o}mez-Mor{\'a}n}, {Schreiber}, {Girven}, \&
  {G{\"a}nsicke}}]{rebassamansergasetal11-1}
{Rebassa-Mansergas}, A., {Nebot G{\'o}mez-Mor{\'a}n}, A., {Schreiber}, M.~R.,
  {Girven}, J., {G{\"a}nsicke}, B.~T., 2011, MNRAS, 413, 1121

\bibitem[{{Ribas}(2006)}]{ribas06-1}
{Ribas}, I., 2006, Ap\&SS, 304, 89

\bibitem[{{Roberts} et~al.(1997){Roberts}, {Gelman}, \&
  {Gilks}}]{robertsetal97-1}
{Roberts}, G.~O., {Gelman}, A., {Gilks}, W.~R., 1997, Annals of Applied
  Probability, 7, 110

\bibitem[{{Schreiber} \& {G{\" a}nsicke}(2003)}]{schreiber+gaensicke03-1}
{Schreiber}, M.~R., {G{\" a}nsicke}, B.~T., 2003, A\&A, 406, 305

\bibitem[{{Schreiber} et~al.(2008){Schreiber}, {G{\"a}nsicke}, {Southworth},
  {Schwope}, \& {Koester}}]{schreiberetal08-1}
{Schreiber}, M.~R., {G{\"a}nsicke}, B.~T., {Southworth}, J., {Schwope}, A.~D.,
  {Koester}, D., 2008, A\&A, 484, 441

\bibitem[{{Southworth} \& {Clausen}(2007)}]{southworth+clausen07-1}
{Southworth}, J., {Clausen}, J.~V., 2007, A\&A, 461, 1077

\bibitem[{{Steele} et~al.(2008){Steele}, {Bates}, {Gibson}, {Keenan},
  {Meaburn}, {Mottram}, {Pollacco}, \& {Todd}}]{steeleetal08-1}
{Steele}, I.~A., {Bates}, S.~D., {Gibson}, N., {Keenan}, F., {Meaburn}, J.,
  {Mottram}, C.~J., {Pollacco}, D., {Todd}, I., 2008, in Society of
  Photo-Optical Instrumentation Engineers (SPIE) Conference Series, vol. 7014
  of \emph{Presented at the Society of Photo-Optical Instrumentation Engineers
  (SPIE) Conference}

\bibitem[{{Steele} et~al.(2004)}]{steeleetal04-1}
{Steele}, I.~A., et~al., 2004, in {J.~M.~Oschmann Jr.}, ed., Society of
  Photo-Optical Instrumentation Engineers (SPIE) Conference Series, vol. 5489
  of \emph{Presented at the Society of Photo-Optical Instrumentation Engineers
  (SPIE) Conference}, p. 679

\bibitem[{{Steinfadt} et~al.(2010){Steinfadt}, {Kaplan}, {Shporer}, {Bildsten},
  \& {Howell}}]{steifadtetal10-1}
{Steinfadt}, J.~D.~R., {Kaplan}, D.~L., {Shporer}, A., {Bildsten}, L.,
  {Howell}, S.~B., 2010, ApJ Lett., 716, L146

\bibitem[{{Tappert} et~al.(2007){Tappert}, {G{\"a}nsicke}, {Schmidtobreick},
  {Aungwerojwit}, {Mennickent}, \& {Koester}}]{tappertetal07-1}
{Tappert}, C., {G{\"a}nsicke}, B.~T., {Schmidtobreick}, L., {Aungwerojwit}, A.,
  {Mennickent}, R.~E., {Koester}, D., 2007, A\&A, 474, 205

\bibitem[{{Tappert} et~al.(2011){Tappert}, {G{\"a}nsicke}, {Schmidtobreick}, \&
  {Ribeiro}}]{tappertetal11-2}
{Tappert}, C., {G{\"a}nsicke}, B.~T., {Schmidtobreick}, L., {Ribeiro}, T.,
  2011, A\&A, in press,arXiv:1107.3586

\bibitem[{{Torres}(2007)}]{torres07-1}
{Torres}, G., 2007, ApJ Lett., 671, L65

\bibitem[{{Townsley} \& {G{\"a}nsicke}(2009)}]{townsley+gaensicke09-1}
{Townsley}, D.~M., {G{\"a}nsicke}, B.~T., 2009, ApJ, 693, 1007

\bibitem[{{Verbunt} \& {Rappaport}(1988)}]{verbunt+rappaport88-1}
{Verbunt}, F., {Rappaport}, S., 1988, ApJ, 332, 193

\bibitem[{{Verbunt} \& {Zwaan}(1981)}]{verbunt+zwaan81-1}
{Verbunt}, F., {Zwaan}, C., 1981, A\&A, 100, L7

\bibitem[{{Webbink}(2008)}]{webbink08-1}
{Webbink}, R.~F., 2008, in {E.~F.~Milone, D.~A.~Leahy, \& D.~W.~Hobill}, ed.,
  Astrophysics and Space Science Library, vol. 352 of \emph{Astrophysics and
  Space Science Library}, p. 233

\bibitem[{{Wood}(1995)}]{wood95-1}
{Wood}, M.~A., 1995, in {D.~Koester \& K.~Werner}, ed., White Dwarfs, vol. 443
  of \emph{Lecture Notes in Physics, Berlin Springer Verlag}, p.~41

\bibitem[{{York} et~al.(2000)}]{yorketal00-1}
{York}, D.~G., et~al., 2000, AJ, 120, 1579

\bibitem[{{Zorotovic} et~al.(2010){Zorotovic}, {Schreiber}, {G{\"a}nsicke}, \&
  {Nebot G{\'o}mez-Mor{\'a}n}}]{zorotovicetal10-1}
{Zorotovic}, M., {Schreiber}, M.~R., {G{\"a}nsicke}, B.~T., {Nebot
  G{\'o}mez-Mor{\'a}n}, A., 2010, A\&A, 520, A86+

\bibitem[{{Zuckerman} et~al.(2003){Zuckerman}, {Koester}, {Reid}, \&
  {H{\"u}nsch}}]{zuckermanetal03-1}
{Zuckerman}, B., {Koester}, D., {Reid}, I.~N., {H{\"u}nsch}, M., 2003, ApJ,
  596, 477

\end{thebibliography}

\bsp

\label{lastpage}

\end{document}